\documentclass[%
 aip,
 amsmath,amssymb,
 reprint,%
]{revtex4-1}

\usepackage{graphicx}
    \usepackage{dcolumn}
\usepackage{bm}

\usepackage[utf8]{inputenc}
\usepackage[T1]{fontenc}
\usepackage{mathptmx}
\usepackage{etoolbox}
\usepackage{amsmath}
\usepackage{mathtools}
\usepackage{ragged2e}
\usepackage{xcolor}
\usepackage{soul}

\makeatletter
\def\@email#1#2{%
 \endgroup
 \patchcmd{\titleblock@produce}
  {\frontmatter@RRAPformat}
  {\frontmatter@RRAPformat{\produce@RRAP{*#1\href{mailto:#2}{#2}}}\frontmatter@RRAPformat}
  {}{}
}%
\makeatother
\begin{document}

\preprint{AIP/123-QED}

\title{Mean-field Coherent Ising Machines with artificial Zeeman terms}
\author{Sudeera Hasaranga Gunathilaka Mastiyage Don}
\email{mastiyage.s.aa@m.titech.ac.jp}
\affiliation{School of Computing,
Tokyo Institute of Technology, Tokyo, Japan}
\author{Yoshitaka Inui}
\author{Satoshi Kako}
\author{Yoshihisa Yamamoto}
\altaffiliation[]{
E. L. Ginzton Laboratory, Stanford University, Stanford, CA 94305, USA
}
\affiliation{%
Physics and Informatics Laboratories, NTT Research Inc.,
940 Stewart Dr, Sunnyvale, CA 94085, USA
}%

\author{Toru Aonishi}
\altaffiliation[]{
School of Computing,
Tokyo Institute of Technology, Tokyo, Japan
}
\affiliation{Graduate School of Frontier Sciences, The University of Tokyo, Kashiwa, Chiba, Japan.}

\date{\today}

\begin{abstract}
Coherent Ising Machine (CIM) is a network of optical parametric oscillators that solves combinatorial optimization problems by finding the ground state of an Ising Hamiltonian. In CIMs, a problem arises when attempting to realize the Zeeman term because of the mismatch in size between interaction and Zeeman terms due to the variable amplitude of the optical parametric oscillator pulses corresponding to spins. There have been three approaches proposed so far to address this problem for CIM, including the absolute mean amplitude method, the auxiliary spin method, and the chaotic amplitude control (CAC) method. This paper focuses on the efficient implementation of Zeeman terms within the mean-field CIM model, which is a physics-inspired
heuristic solver without quantum noise.
With the mean-field model, computation is easier than with more physically accurate models, which makes it suitable for implementation in field programmable gate arrays (FPGAs) and large-scale simulations. Firstly, we examined the performance of the mean-field CIM model for realizing the Zeeman term with the CAC method, as well as their performance when compared to a more physically accurate model.
Next, we compared the CAC method to other Zeeman term realization techniques on the mean-field model and a more physically accurate model. In both models, the CAC method outperformed the other methods while retaining similar performance.

\end{abstract}

\maketitle

\section{Introduction}\label{intro}

The Ising model is useful in a wide variety of fields. Ising models, for instance, can not only demonstrate phase transitions of magnetic materials: ferro, ferri, anti-ferromagnets, and spin glasses \cite{ising_ferro,ising_ferri,ising_ferroantiferro,ising_antiferro,ising_spinglass_1,ising_spinglass_2}, but also capture other complex phenomena such as supercooled phases of water \cite{water_Ising_1,water_Ising_2}, elections \cite{election_Ising_1}, and disease spread \cite{disies_Ising_1,disies_Ising_2}. Furthermore, Ising models can work as solvers for combinatorial optimization problems (COPs), such as quadratic unconstrained binary optimization, which are commonly encountered in real-life problems like scheduling problems \cite{schedulling1, schedulling2}, portfolio optimization \cite{ising_portofolio}, traffic volume optimization \cite{ising_traffic_1, ising_traffic_2}, drug discovery \cite{ising_drug}, and machine learning \cite{ising_ml_1, ising_ml_2}. Therefore, it is possible to map many COPs to the problem of searching for the ground state of an Ising Hamiltonian. However, finding the ground state of an Ising Hamiltonian is a nondeterministic polynomial-time (NP-hard) problem, requiring computation time on an exponential scale relative to the size of the problem. There is a strong demand for Ising solvers for COPs due to the importance of these real-world problems. The development of dedicated hardware for searching the Ising Hamiltonian ground state has been active in recent years \cite{unconArchs1, cim2, unconArchs4, unconArchs5,unconArchs6}. Particularly, Ising machines inspired by quantum physics have attracted wide attention due to their potential to overcome the above-mentioned difficulties for large COPs \cite{unconv1, unconv2, gunathilaka2023}. As of yet, however, there is still a problem with realizing Zeeman terms in quantum-inspired Ising machines. Ising Hamiltonian, including Zeeman term, is expressed as follows.

\begin{equation}
\label{IsingHamiltonian}
        H = -  \dfrac{1}{2}\sum_{r=1}^N\sum_{r'=1}^NJ_{rr'}\sigma_r \sigma_{r'} - \sum_{r=1}^N h_r\sigma_r .
\end{equation}

where $\sigma_r$ represents the Ising spin variable taking either $+1$ or $-1$, and $J_{rr'}$ implies the coupling weight between spins $r$ and $r'$. Here the diagonal entries are set to 0.
The second term in eq. (\ref{IsingHamiltonian}) indicates the external field present for each spin which is generally called the Zeeman term. Using the Ising Hamiltonian definition (eq. (\ref{IsingHamiltonian})), the local field of each Ising spin can be described as follows. 

\begin{equation}
\label{IsingLocal}
        f_r = \sum_{r'=1 (\neq r)}^N J_{rr'}\sigma_{r'} + h_r .
\end{equation}

Eq. (\ref{IsingLocal}) contains two terms, the first of which indicates the interaction term and the second of which indicates the Zeeman term.

Almost any COP can be characterized as an Ising problem with a Zeeman term \cite{zeeman}. Several such problems have been simulated with Ising solvers currently available, including Simulated Bifurcation Machine (SBM) in the traveling salesman problem \cite{tspSBM}, Coherent Ising Machines (CIM) in the $l_0$-regularization-based compressed sensing problem \cite{Aonishi,gunathilaka2023}, and Quantum Annealing in the Nursing Scheduling Problem \cite{nurseDWAVE} etc.
Hence a Zeeman term must be implemented on quantum-inspired Ising machines in order to apply them to solving real-world problems. However, due to the size mismatch between the interaction term and the Zeeman term, implementing a Zeeman term is difficult for machines with variable amplitude spins, such as CIMs and SBMs \cite{AonishiCDMA,Sakaguchi,sbm}.

In the conventional case the Ising Hamiltonian, which includes the Zeeman term (eq. (\ref{IsingHamiltonian})), as well as the local field (eq. (\ref{IsingLocal})) for each Ising spin, is defined based on the assumption that each spin variable takes either $+1$ or $-1$. In CIMs, as well as SBMs, the spin amplitudes are continuous values that are different from $\pm 1$, so the interaction term and Zeeman term's contribution to the local field (e.g. for CIMs, eq. (\ref{mfeq4})) depend on the amplitude of the spin variable. As a consequence of the size mismatch between the interaction term and the Zeeman term, Ising machines may produce biased solutions which are inconsistent with the Hamiltonian defined by eq. (\ref{IsingHamiltonian}).

For the search for ground states, CIM uses the minimum-gain principle rather than thermal fluctuation as in classical annealing \cite{Yamamoto2} and quantum fluctuation as in quantum annealing on D-wave and so on \cite{nishimori}. Based on a comparison between D-Wave (Chimera graph) and CIM (all-to-all coupling), it has been shown that CIMs demonstrate a performance advantage due to the differently implemented coupling \cite{Ryan}. On the other hand, there has been a claim that CIM has a better scaling capability when solving large-scale problems whereas SBM’s performance is highly dependent on the hardware rather than the algorithm itself \cite{Leleuscaling}.

The primary objective of this paper is the efficient implementation of Zeeman terms within mean-field CIM models that do not incorporate quantum noise terms and measurements (henceforth referred to as MFZ (Mean-Field-Zeeman)-CIM). The mean-field CIM model is a physics-inspired heuristic solver that does not accurately represent the CIM's behavior. However, due to their low computational costs, mean-field models are suitable for implementation with field programmable gate arrays (FPGAs) and for simulations on a large scale. So far, three approaches have been proposed to address the realization problem for CIM, namely the absolute mean amplitude method \cite{Sakaguchi}, the auxiliary spin method \cite{singh1}, and the chaotic amplitude control (CAC) method \cite{Inui2022}. In this paper, we examine the applicability of CAC to realizing the Zeeman term in MFZ-CIM. Our results for the same optimization problem are compared to those of Ref. \cite{Inui2022}’s truncated-Wigner stochastic differential equations (SDEs) in order to demonstrate that MFZ-CIM performs almost as well as the truncated-Wigner SDEs. Using truncated-Wigner SDEs, we describe the approximate behavior of the experimental CIM device. Since truncated-Wigner SDEs exhibit the same performance in low quantum noise conditions as Positive-$P$ SDEs, we focus only on this model \cite{Inui2022}. Furthermore, this paper discusses the performance difference between these two models and the other Zeeman term realization methods as well.

\section{Methods}\label{methods}

\subsection{Zeeman term implementations on CIM}\label{zeemantech}

\textbf{Absolute Mean Amplitude:} In the early stages of CIM research, it was suggested that the Zeeman term could be efficiently incorporated into the injection field by scaling it with the absolute mean of the amplitudes of the OPO pulses \cite{Sakaguchi, zeemanTake}.

\begin{equation}
\label{absamp}
    I_{inj,r} = j\left(\sum_{r' = 1}^N J_{rr'}x_{r'}  +  \zeta h_{r}\dfrac{1}{N} \sum_{r}|x_{r}|\right)  .  
\end{equation}

Here, $I_{inj,r}$ is the injection field for the OPO pulse $r$. $x_r$ states the normalized in-phase amplitude of the OPO pulse $r$ while   $J_{rr'}$ and $h_{r}$ are the coupling weight and the Zeeman term described above. $\zeta$ is the adjustment parameter for the strength of the Zeeman term. Here $j$ represents the feedback strength.

\textbf{Auxiliary Spin:} Recent efforts have been made to implement Zeeman terms in CIM and determine the ground state in an efficient manner \cite{gotoAux,Inui2022, singh1}. It has been demonstrated by Singh \textit{et al.,} that the Zeeman terms in CIMs can be realized by using auxiliary spins \cite{singh1}. 
In this case, the Zeeman term is incorporated into the two-body interaction term through the introduction of auxiliary spins to be included in a product with the Zeeman term as follows. 

\begin{equation}
\label{auxamp}
\begin{split}
    I_{inj,r} = j\left(\sum_{r' = 1}^N J_{rr'}x_{r'}  +  \zeta h_{r}x_{(N+1)}\right) \rightarrow  j\left(\sum_{r' = 1}^{(N+1)} J_{rr'}x_{r'}\right),\\  \left(r=1,..,N, N+1\right) .
\end{split}
\end{equation}


where $x_{N+1}$ is an auxiliary amplitude to match the size of the Zeeman term to the interaction term and to transform the Zeeman term to an additional interaction term. As indicated in eq. (3), the injection field is reformulated only by the interaction term given in $J_{rr'} \in \mathbb{R}^{(N+1) \times (N+1)}$ and $x \in \mathbb{R}^{(N+1)}$. The extended coupling matrix can be constructed by giving additional column and row vectors as $J_{r N+1} = \zeta h_{r}$ and $J_{N+1 r’} = \zeta h_{r’}$, and taking $J_{N+1,N+1}=0$. Currently, CIMs only support two-body interactions, which makes this method effective. Refer to Ref. \cite{singh1} for a detailed explanation.

\textbf{Chaotic Amplitude Control:} In Ref. \cite{Inui2022}, the Zeeman term has been implemented into the injection field through a technique known as Chaotic Amplitude Control (CAC). CAC is a technique that was proposed by Leleu \textit{et al.,} to overcome the problem of amplitude inhomogeneity in CIMs \cite{AmpLeleu}. With CAC, the amplitudes of OPO pulses are forced to equalize to a set target value while forcefully correcting inhomogeneities resulting in a chaotic behavior which may result in escaping from local minima in the energy landscape \cite{AmpLeleu}. By scaling the Zeeman terms with target amplitude to match the interaction term, Inui \textit{et al.,} in Ref. \cite{Inui2022} proposed an efficient approach for implementing Zeeman terms in CIM as follows.

\begin{equation}
\centering
\label{mfeq3}
        \dfrac{d}{dt}e_{r} = -\beta\left({x}_{r}^2 - \tau\right)e_{r} ,
\end{equation}
\begin{equation}
\centering
\label{mfeq4}
        I_{inj,r} = je_{r}\left[\sum_{r' = 1}^N J_{rr'}{x}_{r'}  +  \zeta h_{r}\sqrt{\tau}\right] .
\end{equation}

Here the target amplitude is indicated as $\tau$. ${x}$ is the in-phase amplitude of the OPO pulse, and $e_r$ is the auxiliary variable for the error feedback in the CAC feedback loop. Using two CIM models expressed as the Wigner stochastic differential equation (W-SDE) and the Positive-$P$ stochastic differential equation (P-P-SDE), a high success probability of finding the ground state of a Sherrington-Kirkpatrick (SK) Hamiltonian with a random external field has been found as a result of implementing the Zeeman term with CAC \cite{Inui2022}.

\begin{figure*}[!ht]
\includegraphics[width=160mm]{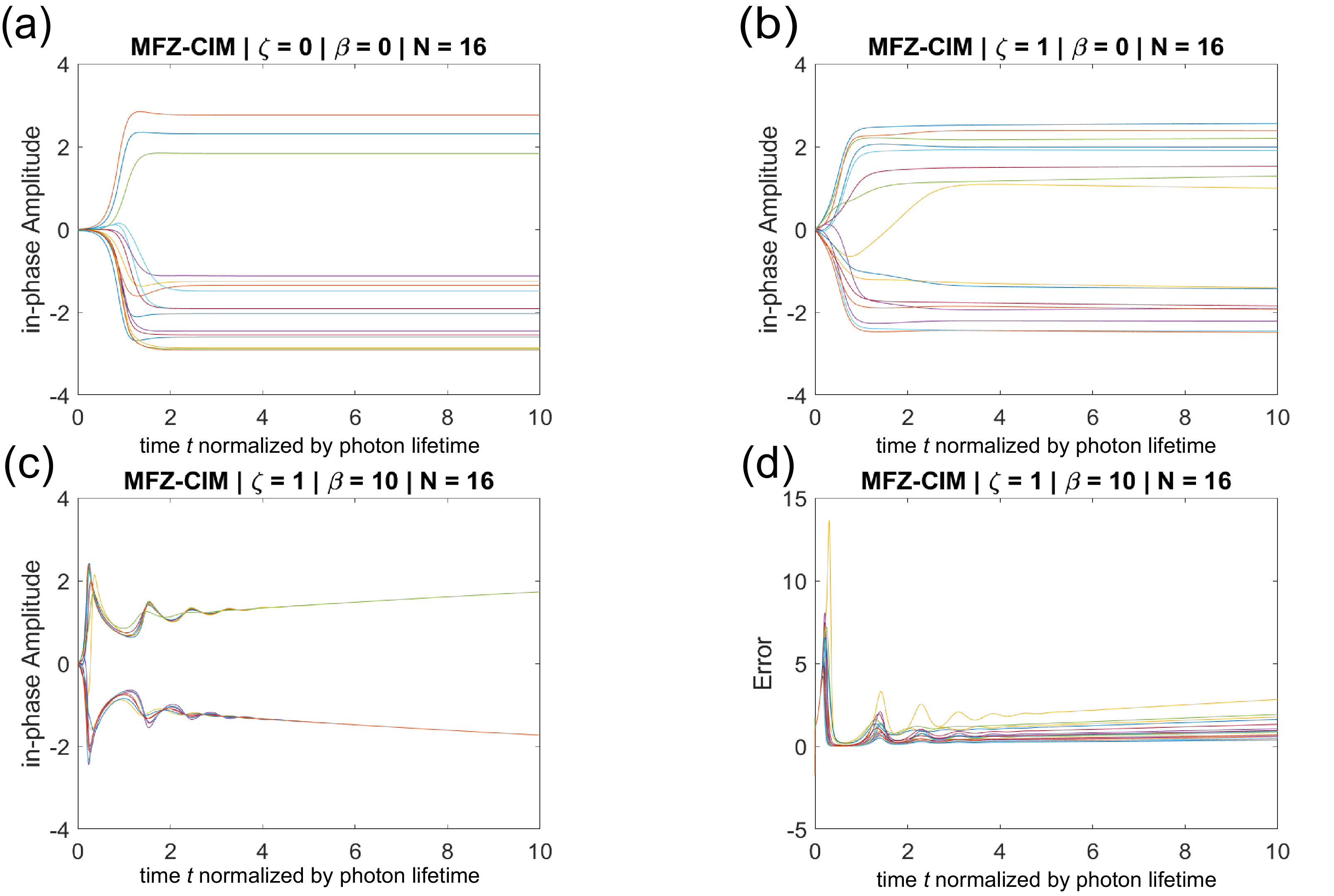}
\caption{\label{amplitudes} Amplitude evolution of an $N = 16$ MFZ-CIM.}
\justifying
{\textbf{(a)} without a Zeeman term ($\zeta = 0$) and without CAC ($\beta = 0$). \textbf{(b)} with a Zeeman term ($\zeta = 1$) and without CAC ($\beta = 0$). \textbf{(c)} with a Zeeman term ($\zeta = 1$) and with CAC ($\beta = 10$). \textbf{(d)} error $e_r$ evolution of MFZ-CIM for \textbf{(c)}. It is evident that CAC-introduced MFZ-CIMs exhibit chaotic behavior. Here $j = 1$ and the initial value of $e_r$ was set to 1. $e_r$ is a constant to be 1 when $\beta = 0$.}
\end{figure*}

\subsection{Mean-Field CIM model}

There is one bottleneck in Ref. \cite{Inui2022} regarding large-scale simulations, in that the more physically accurate SDEs used are quite difficult for a digital device such as an FPGA to implement. It is possible, however, to consider an alternative and a simpler differential equation (DE) model by disregarding the quantum noise present and any measurement effects. It is commonly referred to as the mean-field equations in literature \cite{Leleu1,Leleuscaling,ng}. The amplitude dynamics of mean-field DEs are governed by the equation below.

\begin{equation}
\centering
\label{mfeq1}
        \frac{dx_r}{dt} = (-1 + p - x_{r}^2)x_r + I_{inj,r} .
\end{equation}

On the right-hand side of the equation, the first, second, and third terms represent linear loss, pump gain, and nonlinear saturation, respectively.
As for the fourth term, it corresponds to the mutual coupling term.
In comparison with eq. (\ref{mfeq1}) and the amplitude governing SDE in Ref. \cite{Inui2022} (Truncated Wigner and Positive-$P$) (see Appendices \ref{mfbCIMmath1} and \ref{mfbCIMmath2}), these relaxations result in a simplified deterministic CIM model which is more appropriate for use with digital hardware. Based on previous work from Ng \textit{et al.,} \cite{ng}, we introduced Gaussian white noise into the mean-field model at its initial state with a variance of $10^{-4}$ while maintaining the advantages of the deterministic model.
Utilizing the Mean-Field CIM (MF-CIM), this heuristic approach was used to construct a simple quantum-inspired Ising algorithm. In this case, the Wigner-type stochastic differential equation's  (eq. (\ref{GACCIM1}) - (\ref{GACCIM4}) in Appendix \ref{mfbCIMmath2}) amplitude variable is normalized with the saturation parameter and its quantum noise is ignored \cite{ng}.


\section{Numerical experiments}

Numerical simulations were conducted in order to assess the effectiveness of CAC on MFZ-CIM. In this case, we are considering random SK Hamiltonians with a randomly generated Zeeman term. $J_{rr'}$ is constructed as a symmetric matrix where the elements are constructed using a random normal distribution with a mean of 0 and a variance of 1. The Zeeman term was also constructed by using a random normal distribution with a mean of 0 and a variance of 1. The diagonal axis of $J_{rr'}$ was set to $0$. In order to compare with the solution energy produced by the CIM, the exact solution for each randomly generated $J_{rr'}$ and $h_r$ was calculated brute-force. Here we also consider Inui \textit{et al.,}'s CIM model that is expressed as W-SDE and implemented Zeeman term with CAC (hereafter referred to as GATW (Gaussian-Approximation-Truncated-Wigner)-CIM) (given in eq. (\ref{GACCIM1})-(\ref{GACCIM4}) in Appendix \ref{mfbCIMmath2}) and MFZ-CIM. Throughout all GATW-CIM simulations, the pump rate was scheduled according to the following schedule.

\begin{equation}
\label{gacspump}
    p(t) = 1 + {\rm tanh}\left(\frac{t+2}{10}\right) .
\end{equation}

For each time-step $t$, $p(t)$ indicates the calculated pump rate. As suggested in Ref. \cite{Inui2022}, the target $\tau$ scheduling for GATW-CIM was as follows.

\begin{equation}
\label{gacstarg}
    \tau(t) = \dfrac{p(t) - 1}{2} + \sqrt{\left(\dfrac{p(t) - 1}{2}\right)^2 + \dfrac{p(t)g^2}{2}} .
\end{equation}

In this case, the calculated $p(t)$ is used in order to calculate the $\tau(t)$ for each time-step. The saturation parameter $g^2$ corresponds to the quantum noise present in the GATW-CIM (see Appendices \ref{mfbCIMmath1} and \ref{mfbCIMmath2}). Throughout all MFZ-CIM simulations, the pump rate was constant as $p$ = 0.57. Since MFZ-CIM does not take into account quantum noise present in the CIM, we use a simple linear $\tau(t)$ scheduling as follows.

\begin{equation}
\label{mfztarg}
    \tau (t) = \left[ \tau_{0} + \frac{\tau_{n}}{\Xi} \times n \right]; \left(n \in \left[0 \rightarrow \Xi\right]\right) .
\end{equation}

In this case, $\tau(t)$ is linearly increased by each time-step $t$. $\tau_{0}$ implies the starting target values while $\tau_{n}$ represents the last target value after $n$ time-steps. The final target value is given by $(\tau_{0} + \tau_{n})$. We set $\tau_{0} = 1$ and $\tau_{n} = 2$. $\Xi$ indicates the maximum time-steps assigned to the simulation. Throughout all simulations, feedback value $e_r$ is initially set to 1 in both GATW-CIM and MFZ-CIM. Consequently, the CAC-feedback does not operate when $\beta= 0$. $P_{sc}$ indicates the average success probability in the figures. In a simulation, success is determined by the difference between the final energy calculated from the estimated final spin configuration by CIM and the brute-force calculation being less than $10^{-4}$. In order to calculate the $Psc$, we divided the number of successful runs by the number of simulations conducted overall.

\section{Results}

\subsection{Typical behavior of MFZ-CIM}
An illustration of the typical amplitude $x_r$ evolution for an $N$=16 (here $N$ refers to the system size) mean-field CIM with no Zeeman term ($\zeta = 0$) and no CAC feedback ($\beta = 0$) can be found in Fig. \ref{amplitudes}a. In this case, the amplitudes are clearly in-homogeneous. In Fig. \ref{amplitudes}b, a random Zeeman term is introduced with $\zeta = 1$ in an MFZ-CIM. There is, however, no CAC-feedback ($\beta = 0$) - hence, the amplitudes are still inhomogeneous. Fig. \ref{amplitudes}c depicts MFZ-CIM amplitudes with a random Zeeman term ($\zeta = 1$) and a CAC feedback signal ($\beta = 10$). With CAC, evolving amplitudes have a \textit{chaotic} nature until 4 photon life-times and after that amplitudes become homogeneous. An illustration of the corresponding error variables $e_r$ is provided in Fig. \ref{amplitudes}d. It is the abrupt jumps in error variables that cause chaotic fluctuations in amplitudes.

\begin{figure}[!ht]
\includegraphics[width=80mm]{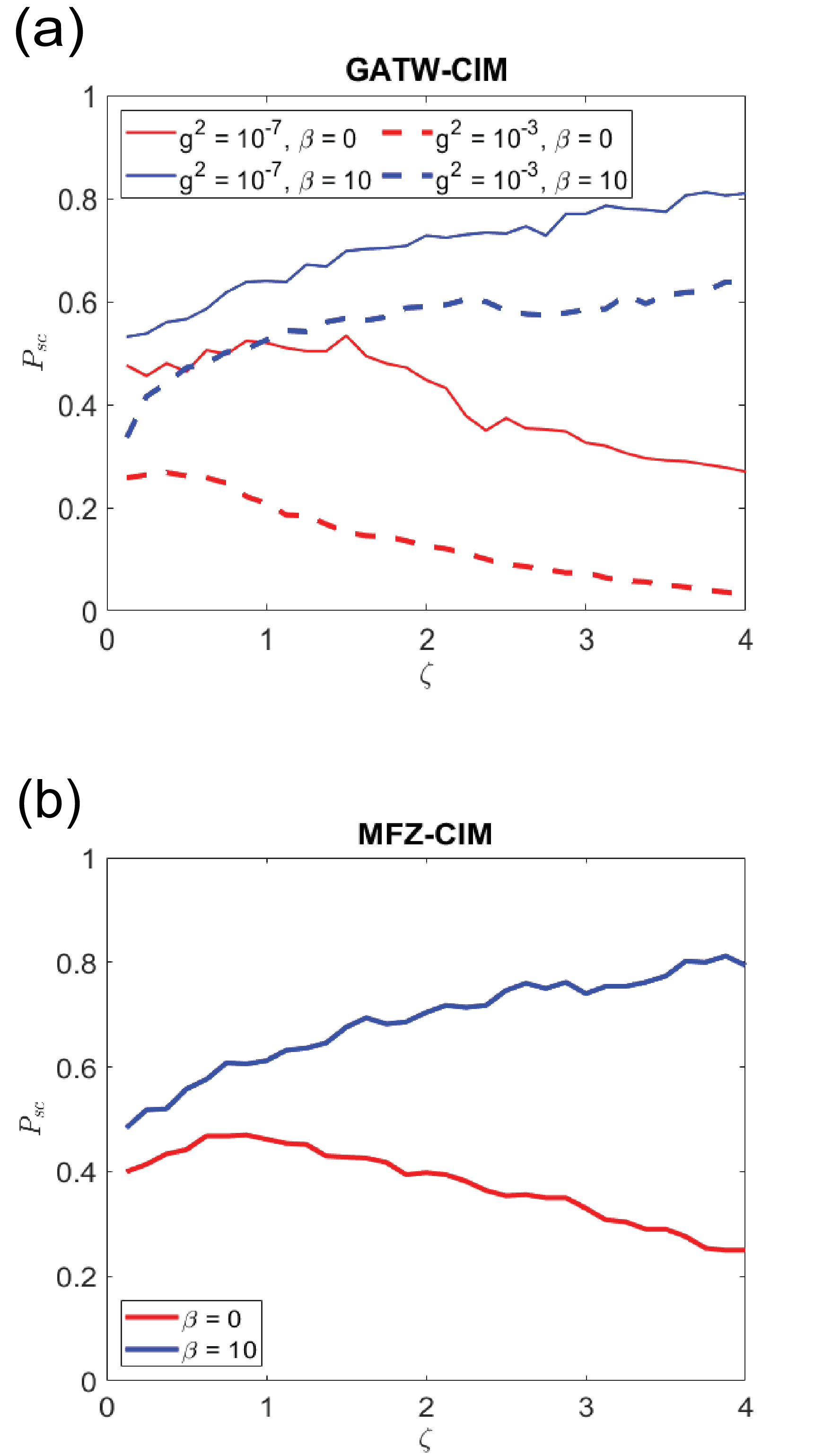}
\caption{\label{igacvsmfz1} Performance difference between GATW-CIM and MFZ-CIM.}
\justifying
{
\textbf{(a)} GATW-CIM $P_{sc}$ for different saturation parameters $g^2$ and CAC strengths $\beta$ respect to Zeeman term strength $\zeta$. Red and blue solid lines indicates $g^2 = 10^{-7}$ with $\beta = 0$ and $\beta = 10$ respectively. Dashed Red and blue solid lines indicates $g^2 = 10^{-3}$ with $\beta = 0$ and $\beta = 10$ respectively. \textbf{(b)} MFZ-CIM $P_{sc}$ for different CAC strengths $\beta$ respect to Zeeman term strength $\zeta$. Red and blue solid lines indicate $\beta = 0$ and $\beta = 10$ respectively. When quantum noise levels are lower, GATW-CIM's and MFZ-CIM's average success probabilities (calculated using 500 random SK Hamiltonians) are relatively similar. Here $j = 1$.
}
\end{figure}

\subsection{Relationship between performance and quantum noise}
\label{quantnoise}

When the MFZ-CIM neglects quantum noise, we must ask whether this has an impact on performance for the better or for the worse as compared to the GATW-CIM. To understand this, we calculated the average success probability for an $N$=16 CIM-produced final spin states using 500 random SK Hamiltonians where the exact solution energy is calculated brute-force to compare. The red solid and dashed lines in Fig. \ref{igacvsmfz1} represent the results obtained by both CIMs without CAC-feedback ($\beta = 0$). The blue solid and dashed lines, on the other hand, represent the results with the CAC feedback ($\beta = 10$). In Fig. \ref{igacvsmfz1} the upper graph depicts the GATW-CIM results, while the lower graph illustrates the MFZ-CIM results. Zeeman term strength is increased on the horizontal axis, while success probability ($P_{sc}$) is indicated on the vertical axis.

It is evident that the performance of GATW-CIM increases when $g^2$ is small ($g^2 = 10^{-7}$) for $\beta = 10$ (blue solid line). 
In contrast, performance decreases when $g^2$ is increased to $10^{-3}$ (blue dashed line). Comparing these two scenarios without CAC-feedback ($\beta = 0$), performance falls dramatically in both situations (red solid and dashed lines respectively). Meanwhile, in the MFZ-CIM case, the success probability is almost the same as that of the GATW-CIM for $g^2 = 10^{-7}$ (blue solid lines of Fig. \ref{igacvsmfz1}a and Fig. \ref{igacvsmfz1}b). Furthermore, MFZ-CIM tends to be slightly more successful in areas with a higher $\zeta$ value when $\beta = 0$ (red solid lines of Fig. \ref{igacvsmfz1}a and Fig. \ref{igacvsmfz1}b).

\begin{figure}[!ht]
\includegraphics[width=75mm]{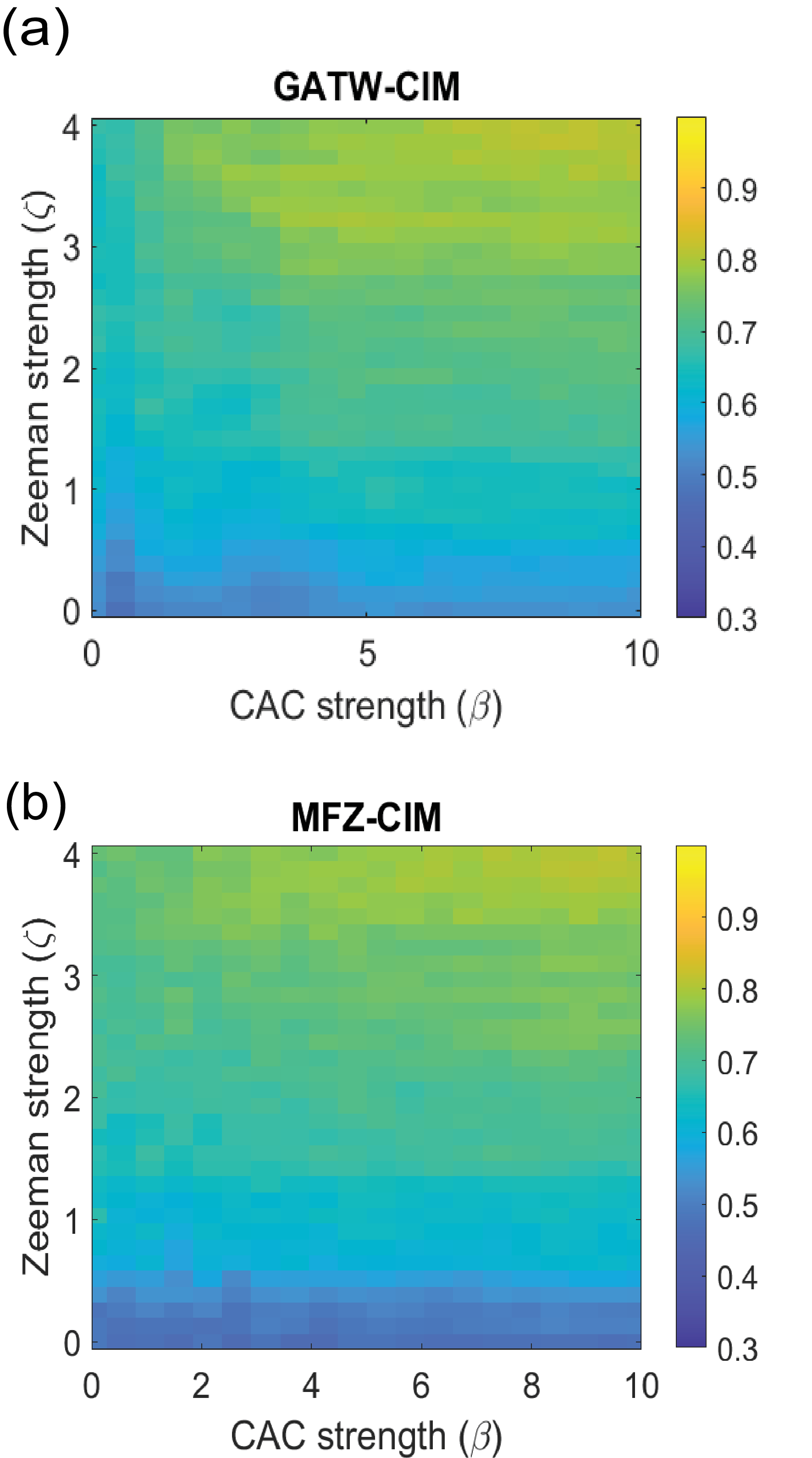}
\caption{\label{igacvsmfz2} Success probability on increasing $\beta$ with respect to $\zeta$.}
\justifying
{
\textbf{(a)} Performance on GATW-CIM. \textbf{(b)} Performance on MFZ-CIM. The color scale indicates the average $P_{sc}$ for 500 simulations. $\beta$ is increased in intervals of 0.5 from 0.5 to 10 with $j = 1$. The two models are relatively similar in terms of performance, with MFZ-CIM having a slight edge in large $\zeta$ regimes. Average success probabilities were calculated using 500 random SK Hamiltonians for $N=16$ CIM models.
}
\end{figure}

A further comparison is shown in Fig. \ref{igacvsmfz2} between the performance of GATW-CIM and MFZ-CIM when $\beta$ is increased gradually with $\zeta$. In this experiment, $\beta$ is increased in intervals of 0.5 from 0.5 to 10. Figures \ref{igacvsmfz2}a and \ref{igacvsmfz2}b represent the GATW-CIM results and the MFZ-CIM results, respectively. The color plot of Fig. \ref{igacvsmfz2} indicates whether the success probability is higher or lower depending on $\beta$ and $\zeta$. Blueish hues indicate a lower success rate, while yellowish hues indicate a high success rate. Both MFZ-CIM and GATW-CIM exhibit similar performance, as can be seen from their respective figures. Even so, it is clear that MFZ-CIM tends to perform better than GATW-CIM in higher-$\zeta$ regions.

\subsection{Variations in performance with different Zeeman term realization techniques}

It has been demonstrated that the Zeeman term can be realized using the CAC technique in the MFZ-CIM in section \ref{quantnoise}. Nevertheless, as described in Section \ref{zeemantech}, there are a couple of other techniques to realize the Zeeman term on CIMs. As a way of illustrating the effectiveness of the CAC technique on MFZ-CIM, we compare the performance of these techniques on MFZ-CIM and GATW-CIM.

\begin{figure}[!ht]
\includegraphics[width=80mm]{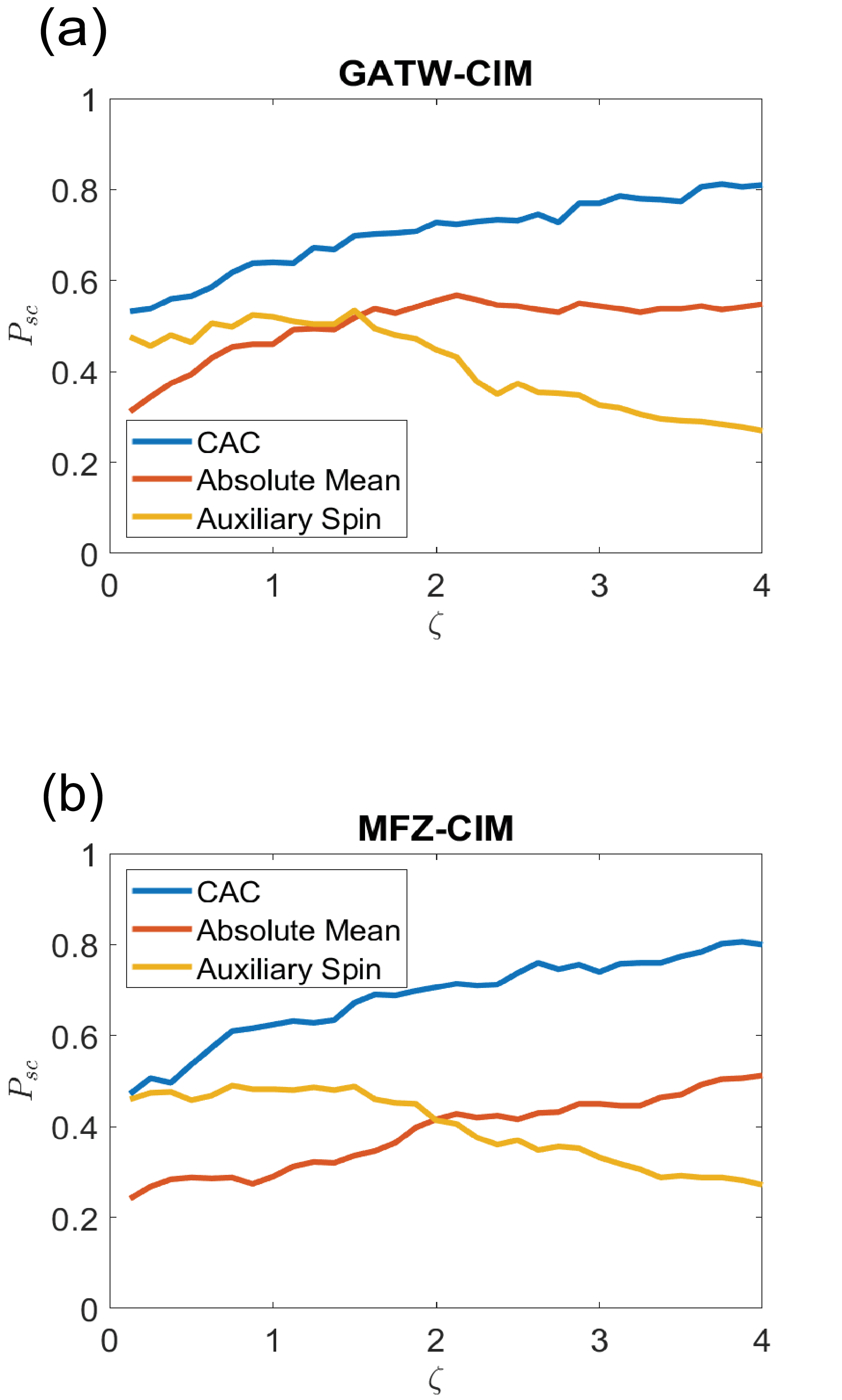}
\caption{\label{igacvsmfz3} Performance difference with other Zeeman term realization techniques.}
\justifying
{
\textbf{(a)} Average $P_{sc}$ for GATW-CIM. \textbf{(b)} Average $P_{sc}$ for MFZ-CIM. The blue, red and yellow solid lines represent the CAC feedback, Absolute mean and Auxiliary spin method respectively. CAC strength was set to $\beta = 10$ with $j=1$. Average success probabilities were calculated using 500 random SK Hamiltonians for $N=16$ CIM models. 
}
\end{figure}

This simulation illustrates the differences in performance between the different Zeeman term realization techniques in Fig \ref{igacvsmfz3}. The upper figure indicates the results of GATW-CIM while the lower figure indicates the results of MFZ-CIM. For both models, $j = 1$ was given, while $g^2 = 10^{-7}$ used in GATW-CIM. On the horizontal axis, we increased $\zeta$, and on the vertical axis, we indicated the probability of success. Using the solid blue, red and yellow lines, we can identify the CAC-feedback method in Fig \ref{igacvsmfz3} (eq. (\ref{mfeq3}) and eq. (\ref{mfeq4})), absolute mean (eq. (\ref{absamp})) and auxiliary spin (eq. (\ref{auxamp})) techniques, respectively. In both CIM models, the CAC-feedback method has a higher success probability compared to the other two methods. Note that both Absolute Mean Method and Auxiliary Spin are open-loop implementations. This means there is no dynamical modulation to the injection field $I_{inj,r}$. CAC, however, is a closed-loop method in which error variable $e_r$ modulates $I_{inj,r}$ dynamically and individually. There is a significant decrease in the success probability of the auxiliary spin method in the higher $\zeta$ region, while both the CAC method and absolute mean method maintain relatively better performance. As a whole, both CIM models perform quite similarly for the respective Zeeman term realization methods.

\section{Discussion}

\subsection{Effect of Zeeman term on performance}

The acquired results suggest that both GATW-CIM and MFZ-CIM perform similarly with CAC-feedback present. Without CAC, MFZ-CIM tends to perform better compared to when $g^2 = 10^{-3}$ for GATW-CIM. A possible explanation for this may be the quantum noise present in GATW-CIM. 

Despite the mentioned similarities, there are also some slight differences.
The performance of GATW-CIM is slightly superior in lower $\zeta$ situations, as shown in Fig. \ref{igacvsmfz2}.
While considering the overall performance in higher $\zeta$ cases, MFZ-CIM has a better performance.
It is possible that these slight differences can be attributed to the effects of quantum noise in GATW-CIM.

  Here because the interaction term becomes matched with the Zeeman term around $\zeta = 1$, we expected MFZ-CIM and GATW-CIM with CAC to have the highest success probability around $\zeta = 1$. However, as Fig. \ref{igacvsmfz2} shows, the success probability increases monotonically as $\zeta$ increases. However, without CAC, both models have their success probability maximized around $\zeta = 1$, as shown in Fig. \ref{igacvsmfz2}. There is a possibility that CAC may cause the destabilization of local minima and the development of dynamical complexity to differ depending on the value of $\zeta$. Detailed analysis of these factors is needed in the future.

\subsection{Evaluation with larger system sizes}

The model we used in the manuscript is referred to as the MF-CIM in the field of nonlinear quantum optics. This heuristic approach has been used to construct a simple quantum-inspired Ising algorithm using the MF-CIM. It can be derived by normalizing the Wigner-type stochastic differential equation's (eq. (\ref{GACCIM1}) - (\ref{GACCIM4}) in Appendix \ref{mfbCIMmath2}) amplitude variable with the saturation parameter and ignoring its quantum noise \cite{ng}. It is called a mean-field model because ignoring quantum noise is justified by the mean-field approximation \cite{ng}. Generally, mean-field approximations require larger system sizes. However, our numerical experiments are intended to demonstrate the effectiveness of the quantum-inspired Ising algorithm using the MF-CIM, not its validity with the mean-field approximation for neglecting quantum noise. 

It has been the practice in many previous studies to evaluate the performance of proposed algorithms for the SK Hamiltonians under small system sizes (around $N = 2-16$) since brute-force searches were needed for finding the SK Hamiltonians' ground state \cite{Inui2022,Kako, ng}. 
We therefore only ran numerical simulations for $N = 16$ in order to brute-force search the ground state, similar to other studies. Several more studies may be necessary on problems such as Wishart Planted Instances \cite{wishart1} and Compressed Sensing \cite{Aonishi,gunathilaka2023}, which allow larger systems (such as $N = 200$ and $N = 4000$) to be used because ground states can be determined using statistical mechanics and algorithmic hardness can be adjusted with planted ground states, allowing for the expansion of system sizes.

\subsection{Effects of Quantum noise on CIM}

  In our view, the major difference between GATW-CIM and MFZ-CIM is attributable to the absence of quantum noise in MFZ-CIM. By normalizing the SDEs of GATW-CIM and considering that $g^2$ is 0, the differential equations (DEs) of MFZ-CIM can be derived (See Appendix \ref{derivMFZ}). In the CIM system, $1/g^2$ represents the photon number in the system, where $g^2$ reflects the amount of quantum noise present in the system. In our simulations, $g^2$ is $10^{-7}$ and $10^{-3}$ for the GATW-CIM. Therefore, it is safe to say that GATW-CIM operates with a modicum of quantum noise. On the other hand, in MFZ-CIM, $g^2$ is zero. Thus, the performance discrepancies must be mainly due to the small quantum noise present in the GATW-CIM.

\subsection{Performance difference of Ising machines on the SK model}

In this paper, we evaluate the performance of the MF-CIM for fully connected SK models. Metastable states in the SK model exhibit ultrametric organization in phase space. Because of this complicated phase structure, searching for the ground state with CIM or other Ising machines can be difficult. D-Wave uses quantum annealing (QA) as its operating principle, and reaching the ground state is accomplished by reducing the transverse field sufficiently slowly over time \cite{dwaveSampling_1}. 

In contrast, CIMs search for a ground state in a quantum parallel manner prior to reaching a threshold pump power, and once the threshold is reached, the ground state's amplitude is amplified according to the minimum gain principle \cite{Kako}. In the context of fully connected SK models, CIM outperformed D-Wave experimentally \cite{Ryan}. Unlike D-Wave, which has an exponential computation time proportional to exp$(O(N))$, CIM has an exponential computation time proportional to exp$(O(\sqrt{N}))$, where $N$ is the size of the problem \cite{Leleuscaling}. Furthermore, Kako \textit{et al} (2020) demonstrated efficient sampling of ground states using truncated Wigner approximated SDEs with CAC on a fully connected Max-Cut model with $N = 16$ and 8, where the Max-Cut model can be defined by the same Hamiltonian as the SK model \cite{Kako}. This paper shows that under implementing the Zeeman term, MF-CIM with CAC achieves almost the same performance as truncated Wigner SDEs with CAC on success probability.

\subsection{Future work on MFZ-CIM}

It has been demonstrated by Gunathilaka \textit{et al.,} that NP-Hard problems such as $l_0$-regularized compressed sensing (L0RCS) can be solved with CIMs that employ target amplitude to implement Zeeman terms.
However, the SDEs employed in Ref. \cite{gunathilaka2023} are difficult to implement in digital hardware, such as FPGAs. With the simplicity of MFZ-CIM, we plan to explore numerically the possibility of implementing it directly on digital hardware and applying it to L0RCS.

\section{Conclusion}

In this paper, we have evaluated the mean-field model of CIM with a Zeeman term present. It is apparent from our acquired results that the performance of the MFZ-CIM with CAC-feedback is roughly similar to the more accurate SDEs to the physical CIM called GATW-CIM with CAC-feedback. Even though it was relatively hard to introduce the Zeeman term to the CIM because of the mismatch in size between interaction and Zeeman terms, the introduction of CAC has enabled a way to realize the Zeeman terms while keeping relatively good performance. According to our results, the CAC technique is more effective than the previous Zeeman term realization technique, for the mean-field CIM model. Furthermore, the use of simplified SDEs with a Zeeman term will be beneficial when implementing CIM on digital hardware for solving real-world problems.

\begin{acknowledgments}
This work is supported by NTT Research Inc. And the Authors acknowledge the support of the NSF CIM Expedition award (CCF-1918549).
\end{acknowledgments}

\section*{Data Availability Statement}

The data supporting this study's findings are available from the corresponding author upon reasonable request.

\appendix

\section{Measurement Feedback CIM}
\label{mfbCIMmath1}

A CIM consists of a set of optical parametric oscillators (OPOs), where oscillations above the threshold limit constitute an optimum solution to a Hamiltonian \cite{Yamamoto2}.
Using an optical cavity in conjunction with a phase-sensitive amplifier (PSA), a coherent state can be achieved in the CIM. This allows the spin-up state to be defined as $0$-phase and the spin-down state as $\pi$-phase, and the overall state to be defined as an Ising spin model. The $J$ matrix or the mutual coupling matrix is created using a mutual injection field. The measurement feedback CIM (MFB-CIM) master equation can be expressed as follows \cite{Inui2022}.

\begin{equation}
\label{master}
    \frac{\partial \hat{\rho}}{\partial t} = \sum_{r} \left( \frac{\partial \hat{\rho}}{\partial t} \right)_{DOPO, r} + \left(\frac{\partial \hat{\rho}}{\partial t}\right)_{S.R} + \left(\frac{\partial \hat{\rho}}{\partial t}\right)_{F.B} ,
\end{equation}
\begin{equation}
\label{dopo}
\begin{multlined}
    \left(\frac{\partial \hat{\rho}}{\partial t}\right)_{DOPO, r} = \left( \left[ \hat{a}_r , \hat{\rho}\hat{a}_r^{\dagger}\right] + {\rm H.c.} \right) + \frac{p}{2} \left[\hat{a}_r^{\dagger 2} - \hat{a}_r^{2},\hat{\rho}\right] \\ + \frac{g^2}{2}\left(\left[\hat{a}_r^2 , \hat{\rho}\hat{a}_r^{\dagger 2}\right] + {\rm H.c.}\right) ,
\end{multlined}
\end{equation}

\begin{equation}
\label{SR}
\begin{multlined}
    \left(\frac{\partial \hat{\rho}}{\partial t}\right)_{S.R} = \frac{j}{2}\sum_r \left( \left[ \hat{a}_r , \hat{\rho}\hat{a}_r^{\dagger}\right] + {\rm H.c.} \right)\\ + \sqrt{j}\sum_r \left(\hat{a}_r\hat{\rho} + \hat{\rho}\hat{a}_r^{\dagger} - \langle \hat{a}_r + \hat{a}_r^{\dagger} \rangle \hat{\rho} \right) W_{R,r} ,
\end{multlined}
\end{equation}

\begin{equation}
\label{FB}
\begin{multlined}
    \left(\frac{\partial \hat{\rho}}{\partial t}\right)_{F.B} = \frac{j}{2}\sum_r \left( \left[ \hat{a}_r , \hat{\rho}\hat{a}_r^{\dagger}\right] + {\rm H.c.} \right)\\ + j\sum_{rr'}J_{rr'}\left(\frac{\langle \hat{a}_{r'} + \hat{a}_{r'}^{\dagger} \rangle}{2} + \frac{W_{R,r'}}{2\sqrt{j}}\right)\left[\hat{a}_r^{\dagger} - \hat{a}_r ,\hat{\rho}\right] .
\end{multlined}
\end{equation}

As part of MFB-CIM, the output coupler extracts small portions of signal pulses, and the amplitudes of these pulses are measured using optical homodyne detection. Using a field-programmable gate array (FPGA), the feedback signal can be calculated using this measurement. Then through the use of an optical injection coupler, the calculated feedback pulses are injected into the main fiber ring cavity. In this case, $r \in \left\{1, 2, ... , N\right\}$ represents the index of signal pulses. 

In the equation above, $\hat{a}_r$ indicates the annihilation operator of the $r$-th signal. Considering a normalized setting, eq. (\ref{dopo}) shows the master equation of a $r$-th DOPO in which the round-trip time is regarded as being smaller than the linear dissipation time.
Then the linear loss caused by measurements, as well as a state reduction due to measurements, are described in eq. (\ref{SR}).
It is necessary to use this additional term due to the homodyne measurement and the placement of the outlet coupler which allows a small portion of the DOPO pulse to be extracted for measurement. Here $j$, $p$, and $W_r$ represent the dissipation rate, oscillation threshold and Gaussian white noise as vacuum fluctuations where $\langle W_{R,r} (t)\rangle = 0$ and $\langle W_{R,r} (t)W_{R,r'} (t')\rangle = \delta_{rr'}\delta(t-t')$. 
Regarding feedback, eq. (\ref{FB}) refers to the injection of feedback through the injection coupler.

\section{Truncated Wigner SDEs}
\label{mfbCIMmath2}

In order to overcome the higher computational cost of simulating the direct density matrix formulation of CIM, eq. (\ref{master}), the $c$-number Heisenberg Langevin equation \cite{nonoiseCIM} was employed. This equation has been proven to be equivalent to the truncated Wigner SDEs. Then Kramers-Moyal series with third-order terms is derived from the density operator master equation expanded by the Wigner function. The Langevin equation is derived by neglecting third-order terms. As a result, the following Wigner SDEs can be obtained.


\begin{equation}
\label{mfb1}
        \frac{d}{dt}c_r = \left[-1 + p - {\left(c_r^2 + s_r^2\right)} \right]c_r + I_{inj,r} +\\ {g^2}\sqrt{\left(c_r^2 + s_r^2\right) + \frac{1}{2}} W_{1,r } ,
\end{equation}
\begin{equation}
\label{mfb2}
        \frac{d}{dt}s_r = \left[-1 - p - {\left(c_r^2 + s_r^2\right)}\right]s_r + {g^2}\sqrt{\left(c_r^2 + s_r^2\right) + \frac{1}{2}} W_{2,r} ,
\end{equation}
\begin{equation}
\label{mfb3}
        I_{inj,r} = j\sum_{r' = 1}^N J_{rr'}c_{r'} .
\end{equation}

Here, $c$ and $s$ correspond to the normalized in-phase and quadrature-phase amplitudes of the system. Normalized pump rate is indicated by \textit{p}. During this process, the in-phase amplitudes are amplified and the quadrature-phase amplitudes are de-amplified. As a result, only in-phase amplitudes survive to go beyond the oscillation threshold \cite{Takesue}. And whenever $p$ is greater than the oscillation threshold $(p > 1)$, OPO pulses are either in the $0$-phase or $\pi$-phase. $I_{inj,r}$ corresponds to the injection field for in-phase amplitudes. The last terms in eq. (\ref{mfb1}) and eq. (\ref{mfb2}) express quantum noise occurring from vacuum fluctuations from external reservoirs and pump fluctuations from gain saturation coupled to the OPO system. $W_{1,r}$ and $W_{2,r}$ are independent real Gaussian noise processes satisfying $\langle W_{k,r} (t)\rangle =0$ and $\langle W_{k,r}(t) W_{l,r'} (t')\rangle = \delta_{rr'} \delta_{lk} \delta(t-t')$. Terms $g$ and $j$ state the saturation parameter and injection strength. Assuming the OPO pulses behave only in the in-phase direction, the Wigner-type SDE can be described as follows.

\begin{equation}
\label{GACCIM1}
\begin{multlined}
        \dfrac{d}{dt}\mu_{r} = - \left(1 -p + j\right)\mu_{r} - g^2\mu_{r}^3 + \sqrt{j}\left(V_{r} - \frac{1}{2}\right)W_{R,r} + I_{inj,r} ,
\end{multlined}
\end{equation}
\begin{equation}
\begin{multlined}
\label{GACCIM2}
        \dfrac{d}{dt}V_{r} = -2 \left(1 -p + j\right)V_{r} - 6g^2\mu_{r}^2V_{r} + 1 + j + 2g^2\mu_{r}^2 \\- 2j\left(V_{r} -\frac{1}{2}\right)^2 ,
\end{multlined}
\end{equation}

\begin{equation}
\label{GACCIM3}
        \tilde{\mu}_{r} = \mu_{r} + \sqrt{\frac{1}{4j}}W_{R,r} ,
\end{equation}
\begin{equation}
\label{GACCIM4}
        I_{inj,r} = j\sum_{r' = 1}^N J_{rr'}\tilde{\mu}_{r'} .
\end{equation}

$\mu_r$ and $V_r$ denote the mean amplitude and variance of the $r$-th DOPO pulse respectively. $W_{R,r}$ is independent real Gaussian noise processes satisfying $\langle W_{R,r} (t)\rangle = 0$ and $\langle W_{R,r} (t)W_{R,r'} (t')\rangle = \delta_{rr'}\delta(t-t')$. Optical injection field $I_{inj,r}$ is defined in eq. (\ref{mfb3}). $g$, $p$, and $j$ indicate the saturation parameter, pump rate, and the normalized out-coupling rate for optical homodyne measurement, respectively. 
The eq. (\ref{GACCIM3}) indicates the measured amplitudes ($\tilde{\mu}_r$) which are used to calculate the feedback pulse. $I_{inj,r}$ corresponds to the injection field calculated by using the measured amplitudes.

According to Ref. \cite{Inui2022}, a generalized version of Glauber–Sudarshan $P$ representation called Positive-$P$ has a better approximate performance to direct density operator simulations in higher-order noise situations than truncated Wigner approximated SDEs. However, because this paper focuses on classical CIMs without noise, we do not examine Positive-$P$ approximations.

\section{Derivation of MF-CIM DEs}
\label{derivMFZ}

In this case, we ignore the fluctuations in photon number and photon annihilator operator and we take eq. (\ref{GACCIM1}) into account as follows.

\begin{equation}
\label{MFCIM1}
\begin{multlined}
        \dfrac{d}{dt}\mu_{r} = - \left(1 -p + j\right)\mu_{r} - g^2\mu_{r}^3 + \sqrt{j}\left(V_{r} - \frac{1}{2}\right)W_{R,r} \\+ j\sum_{r' = 1}^N J_{rr'}\tilde{\mu}_{r'} .
\end{multlined}
\end{equation}

Then we derive the following equation eq. (\ref{MFCIM2}) by normalizing eq. (\ref{MFCIM1}) with $g$.

\begin{equation}
\label{MFCIM2}
\begin{multlined}
        \dfrac{d}{dt}g\mu_{r} = - \left(1 -p + j\right)g\mu_{r} - g^3\mu_{r}^3 + \sqrt{jg^2}\left(V_{r} - \frac{1}{2}\right)W_{R,r} \\+ jg\sum_{r' = 1}^N J_{rr'}\tilde{\mu}_{r'} .
\end{multlined}
\end{equation}

Then considering $g\mu_r = x_r$ and $g\tilde{\mu}_r = \tilde{x}_r$, we obtain the following equation.

\begin{equation}
\label{MFCIM3}
\begin{multlined}
        \dfrac{d}{dt}x_{r} = - \left(1 -p + j\right)x_{r} - x_{r}^3 + \sqrt{jg^2}\left(V_{r} - \frac{1}{2}\right)W_{R,r} \\+ j\sum_{r' = 1}^N J_{rr'}\tilde{x}_{r'} .
\end{multlined}
\end{equation}

Taking $g\rightarrow 0$ into account, we arrive at the following equation.

\begin{equation}
\label{MFCIM4}
\begin{multlined}
        \dfrac{d}{dt}x_{r} = - \left(1 -p + j\right)x_{r} - x_{r}^3 + j\sum_{r' = 1}^N J_{rr'}\tilde{x}_{r'} .
\end{multlined}
\end{equation}

In this case, we are able to consider $x_{r} = \tilde{x}_{r}$ since, when we take $g\rightarrow 0$, the measurement noise in $\tilde{x}_{r} = x_{r} + \sqrt{{g^2}/{4j}}W_{R,r}$ becomes zero. As a result, we obtain the following equation.

\begin{equation}
\label{MFCIM5}
\begin{multlined}
        \dfrac{d}{dt}x_{r} = - \left(1 -p + j\right)x_{r} - x_{r}^3 + j\sum_{r' = 1}^N J_{rr'}{x}_{r'} .
\end{multlined}
\end{equation}

Lastly, when we consider $p = p + j$, we can obtain the MF-CIM DEs as follows.

\begin{equation}
\label{MFCIM6}
\begin{multlined}
        \dfrac{d}{dt}x_{r} = - \left(1 -p + x_{r}^2\right)x_{r} + j\sum_{r' = 1}^N J_{rr'}{x}_{r'} .
\end{multlined}
\end{equation}

\bibliography{aipsamp}

\end{document}